\documentstyle[epsfig,psfig,12pt]{elsart}

\newcommand\be{\begin{equation}}
\newcommand\ee{\end{equation}}
\newcommand\bea{\begin{eqnarray}}
\newcommand\eea{\end{eqnarray}}
\newcommand\ba{\begin{array}}
\newcommand\ea{\end{array}}

\newcommand{\gsim}{\lower.7ex\hbox{$\;\stackrel{\textstyle>}{\sim}\;$}}
\newcommand{\lsim}{\lower.7ex\hbox{$\;\stackrel{\textstyle<}{\sim}\;$}}

\def\beq{\begin{equation}}
\def\eeq{\end{equation}}
\def\bea{\begin{eqnarray}}
\def\eea{\end{eqnarray}}
\def\bq{\begin{quote}}
\def\eq{\end{quote}}

\def\bq{\begin{quote}}
\def\eq{\end{quote}}

\input epsf

\begin{document}

\begin{frontmatter}
  
  \title{Discriminating Neutrino See-Saw Models}
\author{M. Hirsch and S. F. King}
\address{{Department of Physics and Astronomy, University of Southampton,
Southampton, SO17 1BJ, U.K. }}

\maketitle
\begin{abstract}
We consider how well current theories can predict neutrino
mass and mixing parameters, and
construct a statistical discriminator which allows us to compare 
different models to each other.
As an example we consider see-saw models based on family symmetry,
and single right-handed neutrino dominance, 
and compare them to each other and to the case of neutrino 
anarchy with random entries in the neutrino Yukawa and
Majorana mass matrices. 
The predictions depend crucially on the range of the undetermined
coefficients over which we scan, and we speculate 
on how future theories might lead to more precise predictions for 
the coefficients and hence for neutrino observables. 
Our results indicate how accurately neutrino masses and
mixing angles need to be measured by future experiments
in order to discriminate between current models.
\end{abstract}

\end{frontmatter}

{\bf 1} 
Neutrino oscillation physics is one of the most rapidly developing areas 
of particle physics \cite{nu2000}. Current atmospheric neutrino data are 
consistent with $\nu_{\mu}\leftrightarrow \nu_{\tau}$ oscillations 
\cite{SKAtm} with $\sin^22\theta_{23}>0.88$ and  $1.5\times 10^{-3}$ eV$^2$ 
$< |\Delta m_{32}^2| <5\times 10^{-3}$ eV$^2$ (90\% CL) \cite{concha}, 
where $\Delta m_{ij}^2= m_i^2-m_j^2$ in units of eV$^2$. CHOOZ data 
\cite{chooz} limits $\sin^22\theta_{13}<0.1-0.3$ over the SuperKamiokande 
(SK) preferred range of $\Delta m_{32}^2$. The solar neutrino problem can 
be solved by four different combinations of oscillation parameters, three 
of which are based on the Mikheyev-Smirnov-Wolfenstein (MSW) mechanism 
\cite{MSW}, while the fourth involves neutrino oscillations in 
vacuum (VAC). These are the well-known large mixing angle (LMA), the small 
mixing angle (SMA) solutions and a third solution involving a large angle 
but a lower $\Delta m_{12}^2$ (LOW solution). The best-fit points of 
the various solutions are approximately ($\Delta m_{12}^2$, 
$\sin^22\theta_{12}$,name) \cite{concha}: 
($3.3\times 10^{-5}$,$0.78$, LMA MSW);
($5.1\times 10^{-6}$, $0.0027$, SMA MSW);
($10^{-7}$,$0.93$, LOW MSW);
($8 \times 10^{-10}$, $0.93$, VAC).

Recently the SNO collaboration \cite{SNO3} has presented its first data 
on the charged current flux of the solar $^8$B neutrinos, which supports 
both the hypothesis of active neutrino oscillations and the standard solar 
model \cite{BPB}. Furthermore results of recent global analyses, performed 
just days after the SNO result \cite{recent} show that the most favoured 
solutions are either the LMA or the LOW  solutions, with a slight preference 
for the LMA solution. The SMA solution is currently disfavoured at about 
3 $\sigma$, while the VAC regions are disfavoured at various confidence 
levels. 

Because oscillation experiments do not measure absolute values of 
neutrino masses, in the numerical part of this work we will investigate the 
following quantities,
\bea
R\equiv |\Delta m_{21}^2|/|\Delta m_{32}^2| \label{R}\\
s_{C}\equiv 4|U_{e3}|^2(1-|U_{e3}|^2) \label{sc}\\
s_{atm}\equiv 4|U_{{\mu}3}|^2(1-|U_{{\mu}3}|^2) \label{satm}\\
s_{\odot}\equiv 4|U_{e2}|^2|U_{e1}|^2
\label{sodot}
\eea
If $U$ is regarded as the lepton mixing matrix, then, in the limit 
$s_C \ll 1$, to very good approximation $s_{C}\approx \sin^22\theta_{13}$,
$s_{atm}\approx \sin^22\theta_{23}$, $s_{\odot}\approx \sin^22\theta_{12}$.
In the following we will call $\theta_{13}$, $\theta_{23}$, $\theta_{12}$ 
the ``CHOOZ'', atmospheric, solar angle, although strictly speaking 
the various experiments measure $s_{C}$, $s_{atm}$, $s_{\odot}$ in 
the limit $s_C \ll 1$. Note that for the LMA MSW solution $R$ is 
of the order of $R \sim {\cal O}(10^{-2})$.

Following the experimental confirmation of the atmospheric neutrino 
problem by the SK collaboration \cite{SKAtm}, there has been  a flood 
of papers on models of neutrino masses \cite{SelModels}. In order to 
plan future neutrino experiments it would be very helpful to know how 
accurate future measurements would need to be to usefully discriminate
between the different models, and eventually to decide which (if any) of 
all these theoretical attempts is the correct one. In order to address 
this question it is important to understand how well current theories 
predict neutrino masses and mixing angles. In view of this we find it 
rather surprising that although hundreds of different models have been 
proposed there has so far been no attempt to evaluate how well the 
different models actually predict the neutrino parameters. The main 
purpose of this paper is to address this issue. 

In most of the neutrino mass models \cite{SelModels} based on the seesaw 
mechanism \cite{seesaw} one usually assumes that there is some underlying 
broken symmetry which generates a small dimensionless expansion parameter 
$\lambda$. The entries in the neutrino mass matrix then will depend on 
$\lambda$ to some - presumably - high power. Such models are called 
``texture'' models, since they produce approximate zeroes in various 
entries in the mass matrices under consideration (textures). Coefficients 
in front of the expansion parameter, which are not predicted in these 
models, are then usually assumed to be order ${\cal O}(1)$ 
numbers, and their influence on the final result is neglected. In this 
work we are going to investigate the relative importance of the unknown 
order ${\cal O}(1)$ coefficients for the first time systematically. 

Our basic idea is to construct the following simple neutrino mass model 
discriminator which we define qualitatively in the following sense: 
calculate the leading structure (given as powers of $\lambda$ which are 
fixed by symmetry) of any interesting model and scan 
``randomly'' over the unknown parameters. From these samples one can 
construct distributions of measurable neutrino parameters, which can be 
compared to experiment. A model is then considered ``good'' if the peaks 
of the distributions coincide with (or are close to) the experimentally 
preferred value. Such a discriminator is by its nature qualitative rather 
then quantitative, and it is unclear how much weight should be attributed
to a model that predicts all quantities close to the experimentally preferred 
values. The main problem of the approach is that we have treated all the 
coefficients on an equal footing, and one can imagine some extra condition 
which could become known in the future which could single out a particular 
value for a ``random'' parameter, which would lead to predictions in the 
tails of the distributions, rather than near the peaks. However, we would 
like to argue that although {\it a priori} such a possibility cannot be 
strictly excluded, it might be considered to be contrary to the spirit of 
texture models.

In our analysis we only consider ``high energy scale'' models based on the 
seesaw mechanism \cite{seesaw}, which are otherwise hard to test. From 
the ``testability'' point of view, low-energy scale models of neutrino 
mass (see, for example \cite{NtrlDecay,Raidal}) might be considered 
preferable. However, even if one of the low-energy approaches is the true 
neutrino mass model, a confirmation certainly lies a number of years in 
the future, and so our methods of discrimination could be useful even in 
that case.

{\bf 2} For definiteness we consider \cite{flavour} models based 
on a $U(1)$ family symmetry. 
The idea of such a symmetry is that the three families of leptons
are assigned different $U(1)$ charges, and these different charges
then control the degree of suppression of the operators responsible
for the Yukawa couplings, leading to Yukawa matrices with 
a hiearchy of entries, and approximate ``texture'' zeroes.
As usual it is assumed that the $U(1)$ 
is broken by the VEVs of some fields $\theta , \bar{\theta}$ 
which are singlets under the standard model gauge group, but
which have vector-like charges $\pm 1$ under the $U(1)$ family
symmetry. The $U(1)$ breaking 
scale is set by $<\theta >=<\bar{\theta} >$. Additional exotic vector 
matter with mass $M_V$ allows an expansion parameter $\lambda$ to be 
generated by a Froggatt-Nielsen mechanism,
\begin{equation}
\frac{<\theta >}{M_V}=\frac{<\bar{\theta} >}{M_V}= \lambda \approx 0.22
\label{expansion}
\end{equation}
where the numerical value of $\lambda$ is motivated by the size of 
the Cabibbo angle. Small Yukawa couplings are generated effectively 
from higher dimension non-renormalisable operators corresponding 
to insertions of $\theta$ and $\bar{\theta}$ fields and hence 
to powers of the expansion parameter in Eq.\ref{expansion}. The number 
of powers of the expansion parameter is controlled by the $U(1)$ charge 
of the particular operator. The lepton doublets, neutrino singlets, 
Higgs doublet and Higgs singlet relevant to the construction of 
neutrino mass matrices are assigned $U(1)$ charges $l_i$, $n_p$, 
$h_u=0$ and $\sigma$. From Eq.\ref{expansion}, the heavy Majorana 
matrix elements are $M_{RR}^{ij} \sim A_{ij}\lambda^{|n_i+n_j+\sigma|}$. 
The neutrino Dirac matrix elements are
$m_{LR}^{ij} \sim a_{ij}\lambda^{|l_i+n_j|}$,
where $A_{ij}$ and $a_{ij}$ are undetermined coefficients.
For details see \cite{Steve0}. 
 
The theory does not determine the coefficients 
$A_{ij}$ and $a_{ij}$ 
which we take to be free parameters 
which are chosen randomly over some range.
The physical light effective Majorana 
neutrino mass matrix which results from the see-saw mechanism \cite{seesaw} 
is given by $m_{LL}=m_{LR}M_{RR}^{-1}m_{LR}^T$, and $m_{LL}$ is 
diagonalised by a matrix $U$ and the eigenvalues $m_i$ are the light 
neutrino masses. In the basis in which the charged lepton
mass matrix is diagonal, $U$ is just the lepton mixing matrix.

\begin{table}
 \begin{center}
  \begin{tabular}{||l|r|r|r|r|r|r|r||} 
\hline
Models & $l_1$ & $l_2$ & $l_3$ & $n_1$ & $n_2$ & $n_3$ & $\sigma$ \cr
\hline
FC1 &  -2 & 0 & 0 & -2 & 1 & 0 & 0 \cr
\hline
FC2 &  -3 & -1 & -1 & -3 & 0 & -1 & 3 \cr
\hline
FC3 &  -1 & 1 & 1 & $\frac{1}{2}$ & 0 & -$\frac{1}{2}$ & -1 \cr
\hline
FC4 &  -1 & 1 & 1 & $\frac{1}{2}$ & -$\frac{1}{2}$ & -$\frac{1}{2}$ & -1 \cr
\hline
\end{tabular}
\end{center}
\caption{Flavour charges (FC) for four models, as discussed
in the text.}
\label{FC}
\end{table}

In Table \ref{FC} we give four examples of models based on different 
choices of flavour charges.  Models FC1-FC3 are examples of single right 
handed neutrino dominance (SRHND) \cite{Steve0} which naturally lead to 
small values of $R$, typically $R\sim \lambda^4$ (see below). The 
essential idea of SRHND is that one of the right-handed neutrinos 
contributes dominantly to the 23 block of $m_{LL}$. This results in 
the approximate vanishing of the the 23 subdeterminant of $m_{LL}$, and 
hence one expects a hierarchical spectrum of masses, 
$m_{\nu_2} \ll m_{\nu_3}$. Without SRHND a large mixing angle 
$\theta_{23} \sim \pi/4$ would typically result in two eigenvalues of the
same order of magnitude $m_2 \sim m_3$, and the presence of
a mass hierarchy could only be achieved at the expense of
some fine-tuning. The amount of fine-tuning required depends
on the value of the mass hierarchy.

FC1 and FC3 are taken from \cite{Steve} where it is shown that they lead 
to the expectation of a large solar angle. FC1 leads to dominant (from 
the third right-handed neutrino) and subdominant contributions to 
$m_{LL}$ \cite{Steve} (order ${\cal O}(1)$ coefficients suppressed 
for simplicity):
\beq
m_{LL}^{FC1}
\sim 
\left( \begin{array}{ccc}
\lambda^4 & \lambda^2 & \lambda^2    \\
\lambda^2 & 1 & 1    \\
\lambda^2 & 1 & 1   
\end{array}
\right)
+ {\cal O}
\left( \begin{array}{ccc}
\lambda^4 & \lambda^2 & \lambda^2    \\
\lambda^2 & \lambda^2 & \lambda^2    \\
\lambda^2 & \lambda^2 & \lambda^2   
\end{array}
\right).
\label{SRHND}
\eeq
We constructed FC2 as an example of a flavour model 
predicting a small solar angle. FC2 gives dominant
and subdominant contributions:
\beq
m_{LL}^{FC2}
\sim 
\left( \begin{array}{ccc}
\lambda^7 & \lambda^5 & \lambda^5    \\
\lambda^5 & \lambda^3 & \lambda^3   \\
\lambda^5 & \lambda^3 & \lambda^3   
\end{array}
\right)
+ {\cal O}
\left( \begin{array}{ccc}
\lambda^9 & \lambda^7 & \lambda^7    \\
\lambda^7 & \lambda^5 & \lambda^5    \\
\lambda^7 & \lambda^5 & \lambda^5   
\end{array}
\right).
\label{SRHND2}
\eeq
FC3 gives a large solar angle, and also a larger CHOOZ angle:
\beq
m_{LL}^{FC3}
\sim 
\left( \begin{array}{ccc}
\lambda & 1 & 1    \\
1 & \lambda^{-1} & \lambda^{-1}    \\
1 & \lambda^{-1} & \lambda^{-1}   
\end{array}
\right)
+ {\cal O}
\left( \begin{array}{ccc}
\lambda & \lambda & \lambda    \\
\lambda & \lambda & \lambda    \\
\lambda & \lambda & \lambda   
\end{array}
\right).
\label{SRHND3}
\eeq
The model FC4 is deliberately
chosen not to have SRHND, but is designed to give the same leading 
order structure as $m_{LL}^{FC3}$, but with subleading contributions
to the 23 block which are also of the same order, leading to 
an unsuppressed value of $R$. We will not consider renormalization 
group effects here, since it has been shown that they only lead to 
corrections of several percent \cite{KingNimai} which are negligible 
compared to the effects of the unknown coefficients. 

Approximate expectations for the 
experimentally accessible quantities ($\theta_{23}$, $\theta_{13}$, 
$\theta_{12}$ and $R\equiv |\Delta m_{12}^2|/|\Delta m_{23}^2|$)
may be estimated from the form of the matrices 
in Eqs.\ref{SRHND}-\ref{SRHND3} as discussed in \cite{Steve} and
are given in Table \ref{FCexp}.

\begin{table}
 \begin{center}
  \begin{tabular}{||l|r|r|r|r||} 
\hline
Model & $\theta_{23}$ & $\theta_{13}$ & $\theta_{12}$ & $R$ \cr
\hline
FC1 & 1 & $\lambda^2$ & 1 & $\lambda^4$ \cr
\hline
FC2 & 1 & $\lambda^2$ &$\lambda^2$  & $\lambda^4$ \cr
\hline
FC3 & 1 & $\lambda$ & 1 & $\lambda^4$ \cr
\hline
FC4 & 1 & $\lambda$ & - & - \cr
\hline
\end{tabular}
\end{center}
\caption{Approximate expectations for 
$\theta_{23}$, $\theta_{13}$, $\theta_{12}$ and $R$ for the 
four different choices of flavour charges defined 
in the text.}
\label{FCexp}
\end{table}

Finally, as a fifth example we will consider neutrino {\em anarchy} 
\cite{HMW}. In the language of flavour charges anarchy simply 
corresponds to choosing $l_1=l_2=l_3$ and $n_1=n_2=n_3$, such 
that the effective neutrino mass matrix shows no obvious structure. 
It is therefore completely determined by the random coefficients.

{\bf 3} As in \cite{HMW} we investigate the quantities
$R$, $s_{C}$, $s_{atm}$ and $s_{\odot}$, as defined in 
Eqs.\ref{R}-\ref{sodot}. 
As mentioned in the introduction these are appropriate quantities when 
considering data from oscillation experiments. Recall, in the limit 
$s_C \ll 1$, to very good approximation $s_{C}\approx \sin^22\theta_{13}$,
$s_{atm}\approx \sin^22\theta_{23}$, $s_{\odot}\approx \sin^22\theta_{12}$.
As a test of our numerical procedure we ran a sample calculation with one 
million randomly chosen points for all five models imposing the same cuts 
as used in \cite{HMW} ($R<1/10$, $s_{C}<0.15$, $s_{atm}>0.5$ and 
$s_{\odot}>0.5$). For this test we restricted the range of the coefficients 
to be in the interval [-1,1] in order to be able to compare our calculation 
to the results presented in \cite{HMW}. Note, however, that this range 
is not the appropriate one for texture models, as discussed below, and 
was used only in this test calculation.
We have checked that our results for the anarchy matrix are 
statistically consistent with the original anarchy calculation in \cite{HMW}, 
namely that approximately 3\% of the points pass all the cuts. By comparison, 
with the same procedure applied to model FC1 in Table \ref{FC} 25\% of all 
points pass the cuts.

In texture models, however, it is necessary to choose the range of the
coefficients $A_{ij}$ and $a_{ij}$ differently from [-1,1]. As explained 
in more detail in section 4, where the importance of the coefficients 
is discussed, in order to not destroy the texture structure 
imposed by the expansion parameter, we have chosen the range of the 
coefficients to be $\pm [\sqrt{2\lambda},1/\sqrt{2\lambda}]$. In other 
words the magnitude of each of the coefficients is selected randomly in 
the range 0.663-1.508 and its sign is also selected randomly.  
With the structure preserving range only 1.8\% of the 
neutrino anarchy points pass the cuts, while for the case with structure 
FC1  56\% of the points now pass the cuts. Structure in the neutrino 
mass matrix is clearly preferred over anarchy, since more points pass the 
cut in that case.

\begin{figure*}
\begin{picture}(0,290)
\put(-40,-250)
{\mbox{\psfig{file=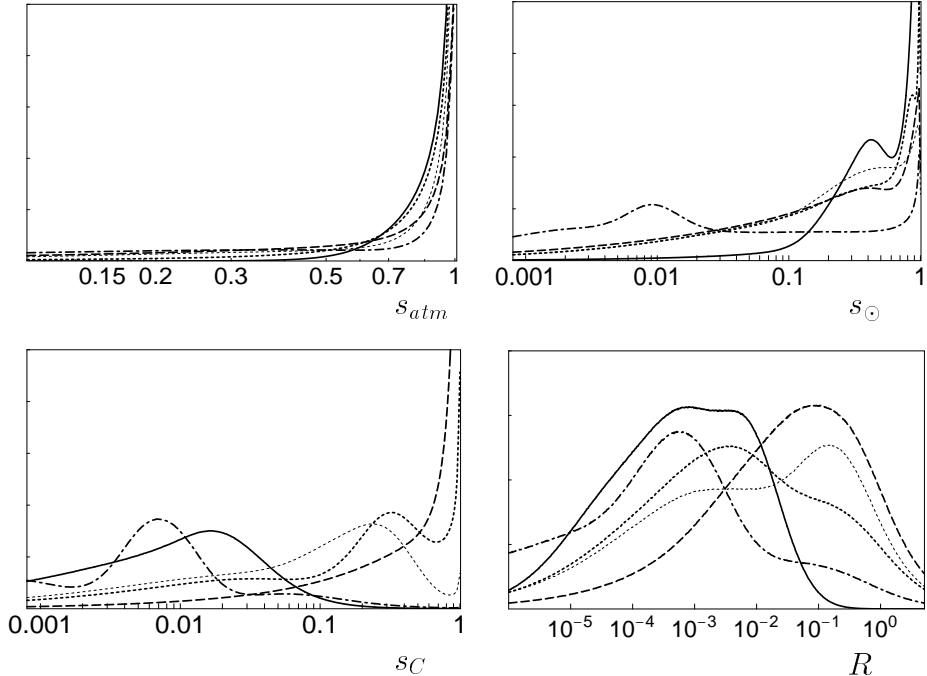,width=1.1\textwidth,
height=0.65\textheight,angle=0}}}
\end{picture}
\caption[allfig]{Theoretical distributions for the predictions of neutrino 
mass and mixing parameters for five selected see-saw models:
FC1 (full), FC2 (dot-dashes), FC3 (thick dots), FC4 (thin dots),
anarchy (dashes). Matrix
coefficients are randomly chosen in the interval 
$\pm [\sqrt{2\lambda},1/\sqrt{2\lambda}]$.
The vertical axis in each panel (deliberately not labelled)
represents the {\em logarithmically binned} 
distributions with correct relative normalisation
for each model, with heights plotted on a linear scale in arbitrary units.}
\label{fig:allplots}
\end{figure*}

In Figure 1 we give distributions for $s_{atm}$, $s_{\odot}$,$s_{C}$ and 
$R$ for $10^8$ points and five different models, using random 
coefficients in the range $\pm [\sqrt{2\lambda},1/\sqrt{2\lambda}]$. 
Note, that the plots are on a log-linear scale. Using a semilogarithmic 
scale is suitable for quantities which a priori can be distributed in 
intervals ranging over several orders of magnitude, as can be expected 
for the observables we are interested in. Choosing a linear scale a 
model predicting $s_{\odot}=0.1$, for example, could not be distinguished 
from a model predicting $s_{\odot}=10^{-3}$. Thus our choice of scale 
and our definition of a ``good'' model corresponds to a rather mild 
requirement. Instead of asking for an acceptance of a given model at 
some (high) level of confidence, we ask for the {\it minimum requirement} 
that models should give at least the correct order-of-magnitude 
estimate for observables. Any model failing even this mild test will 
fail even more badly once subjected to a more sophisticated numerical 
analysis.

The dashed line and the full line correspond to the case of neutrino 
anarchy and the FC1 model as in Table \ref{FC}. If we focus only on 
these to begin with we see that they both give a peak at $s_{atm}=1$, 
and $s_{\odot}=1$, but the FC1 model gives also a peak at 
$s_{\odot}=0.4$, and disfavours $s_{\odot}<0.1$. 

The main difference between anarchy and FC1 is in the 
CHOOZ angle where anarchy again gives a peak at $s_C=1$, while the 
FC1 model has a peak near the small value $s_C=0.02$. The R value 
for the FC1 model also prefers smaller values than the case of 
anarchy, albeit with a broader peak, and may be consistent with 
either the LMA or the LOW solutions. Clearly more accurate experimental 
determinations of physical parameters will eventually provide a 
powerful discriminator between the case of neutrino anarchy and 
neutrino structure.

The remaining lines in Figure 1 correspond to the different flavour 
models FC2, FC3, FC4. All the models predict large values of $s_{atm}$ 
and so cannot be distinguished this way. The FC2 model (dot-dash) gives 
a peak at $s_{\odot}=0.01$, and therefore prefers the SMA MSW solution, 
consistent with the analytic expectations discussed previously \cite{Steve}.
It also has peaks near $s_{C}=0.01$ and $R=10^{-3}$ consistent 
with the analytic expectations. The FC3 model (thick dots) is consistent 
with the LMA or LOW solutions, similar to the FC1 model (solid), since 
it peaks at large $s_{\odot}$ and the $R$ peak is in the same region, 
albeit with a different detailed shape. However, unlike FC1, the FC3 
model has a peak at the larger value $s_{C}=0.3$ and so may be 
distinguished by a measurement of $\theta_{13}$. Finally the FC4 model 
(thin dots) is designed to give the same leading order contributions 
to $m_{LL}$ as FC3 (thick dots), differing only in the 
subleading contributions, 
so that FC4 does not possess the feature of SRHND. This subtle 
difference between the two models gives rise to a quantitative difference
since the peak in $R$ for the FC4 model is shifted to the right 
compared to that in the FC3 model. 

{\bf 4} 
We emphasise that in all flavour models of the kind we consider \cite{flavour}
the coefficients $A_{ij}$ and $a_{ij}$ are not specified. In order for 
these models to make sense it is implicitly assumed that the coefficients 
are of order unity to within an error determined by the expansion parameter.
The range of coefficients applied previously, namely 
$\pm [\sqrt{2\lambda},1/\sqrt{2\lambda}]$, 
was chosen in order that the structure imposed by the flavour charges and 
the expansion parameter $\lambda$ was not destroyed. We re-emphasise that 
this is not guaranteed, but a minimum requirement for texture models in 
order to make sense physically. We think that future theoretical progress
in flavour models of neutrino masses lies in being able to predict these 
coefficients more accurately, and Figure 2 demonstrates the importance 
of this point.

In Figure 2 we have plotted the distribution in the solar angle for the 
case of FC2, but with 3 different ranges of the coefficients. 
As the range of the coefficients is reduced, the central peak
of the distribution becomes more pronounced, with smaller side-peaks
emerging as a result of accidental cancellations due to the arbitrary
signs of each element which we allow in our Monte Carlo.
Thus Figure 2 shows the accuracy of the prediction 
for the solar angle which may be possible in some future
theory where the coefficients are specified more accurately.
Analagous results are obtained for the other observables.

\begin{figure}
\setlength{\unitlength}{1mm}
\begin{picture}(0,55)
\put(0,-20)
{\mbox{\epsfig{file=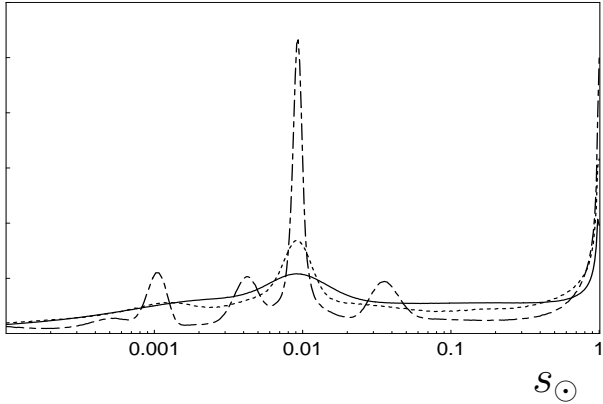,height=10.0cm,width=8.0cm}}}
\end{picture}
{\vskip-7mm\hskip70mm {\Large $s_\odot$}\vskip2mm}
\caption[allfig]{Plots of $s_{\odot}$ for the case of FC2 with 
three different ranges of the coefficients $A_{ij}$ and $a_{ij}$. 
Full line: $\pm [\sqrt{2\lambda},1/\sqrt{2\lambda}]$. Dotted 
line: $\pm [0.82,1.18]$. Dash-dotted line: $\pm [0.95,1.05]$. }
        \label{fig:coeff}
\end{figure}

{\bf 5} 
To conclude, 
we have considered how well current theories, in particular those
based on the see-saw mechanism and SRHND,
can predict neutrino
mass and mixing parameters, and
have proposed a qualitative discriminator based on logarithmic distributions 
calculated from a random scan over the (presently unknown) coefficients in 
the neutrino Yukawa matrix and the neutrino Majorana 
matrix, leading to the distributions
in Figure 1. This paper is the first serious attempt
to show how the uncertainty of the coefficients of the
input see-saw matrices feeds through to the uncertainty in the
predictions of experimental neutrino observables.

What is the value of such distributions, and what conclusions
can be drawn from them? Consider for example the
distributions in the CHOOZ angle $s_C$ corresponding to the
third panel of Figure 1. It is clearly seen that
three of the models FC3 (thick dots), FC4 (thin dots) and
anarchy (dashes) give distributions which peak
at values of $s_C$ which are larger than or around the experimental
limit $s_C\leq 0.1-0.3$. Indeed anarchy peaks at $s_C=1$. 
This panel also shows qualitatively how increasingly unlikely these
models will become as the experimental limit on $s_C$ becomes
stronger, and again anarchy fares the worst having the
lowest number of points in the tail of the distribution.
We should emphasise that such a plot of the distribution in $s_C$
was not shown for anarchy in \cite{HMW}.
On the other hand the models FC1 (full) and FC2 (dot-dash) 
give distributions which peak at much smaller values
of $s_C$, in agreement with the analytic estimates
in Table 2. The value of the distributions is that
one can see pictorially how much a given experimental
limit on $s_C$ favours or disfavours a particular model.
The conclusions which one draws from such distributions
remain a matter of taste, but at least the distributions
enable such conclusions to be made based on the best possible
information. It is manifestly clear that seeing a probability
distribution is more informative than simply having an analytic
estimate of the position of the peak of the distribution,
and our paper supplies this additional information.

Clearly from the above discussion an accurate measurement of $\theta_{13}$ 
is very important, since it could rule out or disfavour large classes of 
models, and this may require a neutrino factory \cite{nf}. What about the 
other neutrino observables? From the first panel in Figure 1 we see that
an accurate determination of the atmospheric angle $\theta_{23}$ is not 
so important since all the considered models peak at maximal mixing. 
However, by contrast, a measurement of the solar angle $\theta_{12}$ is 
important in resolving large classes of models, and the second panel in 
Figure 1 shows the distributions in $s_{\odot}$. The recent SNO indication 
of a large solar angle clearly disfavours the model FC2 (dot-dash) which 
peaks at a small solar angle according to the analytic estimate in Table 2. 
One may ask, what do we learn from the distributions that we do not already 
know from the analytic estimates of the position of the peak? The answer 
is that the distributions give us a rough idea of the uncertainties of 
the analytic estimates.

The distributions in $R$, corresponding to the fourth panel in Figure 1, 
all have rather broad distributions which could not have been predicted by 
the analytic results, which again only predict the position of the peaks. 
One would expect that a small measured value of $R$ would favour SRHND 
over other models where $R$ is larger. This effect is clearly seen since 
the FC4 models (thin dots) and the anarchy model (dashes) do not have 
SRHND and therefore give distributions in $R$ in the fourth panel in 
Figure 1 which are peaked towards larger values than the SRHND models
FC1-FC3. However what could not have been predicted is the width of the 
distributions which are so broad as to make these distributions overlap
over large regions of $R$. Once again the distributions provide this 
additional information which has hitherto been missing from the literature.
Moreover, as seen from Table 2, for the models FC1-FC3 one expects 
$R$ to be $\sim \lambda^4$ in all cases, whereas Fig. 1 demonstrates that 
even the position of the peaks in this variable can vary up to an 
order of magnitude. Again, the distributions give us an estimate of 
the limitations of the analytic approaches. 

We hope that the additional information provided by the distributions
will be of use to our experimental colleagues in planning the next 
generation of neutrino experiments, but we would caution that future 
theories could predict smaller ranges of the unknown coefficients, which 
could demand higher experimental accuracy in the measurement of neutrino 
mass and mixing parameters than that deduced from the results presented here.
Also we emphasise that we have treated all the coefficients as independent 
and uncorrelated, and some future theory could relate these coefficients 
in such a way as to give rise to predictions in the tails of the 
distributions rather than near the peaks. To some extent this undermines 
the conclusions which we can draw from our current analysis. However we 
feel that the distributions which  we present are a fair reflection of 
current theoretical uncertainties.

\bigskip
{\small S.K. is grateful to PPARC for the
support of a Senior Fellowship. M.H. would like to thank E. Nardi 
for discussions and the Universidad de Antioquia for the kind 
hospitality during the early stages of this work. We also would like 
to thank J. Parry and D. Crooks for a careful reading of the 
manuscript.}


\begin{thebibliography}{99}
\bibitem{nu2000} 
See, for example, talks at XIX. Int. Conf. on Neutrino Physics 
\& Astrophysics (Neutrino 2000), 
http://nu2000.sno.laurentian.ca/; A summary of links on pages 
related to neutrino oscillations can be found at 
http://www.hep.anl.gov/ndk/hypertext/nuindustry.html


\bibitem{SKAtm} Y. Fukuda {\it et al.}  [Super-Kamiokande Collaboration],
Phys. Rev. Lett. {\bf 81}, 1562 (1998).

\bibitem{concha} See e.g. 
M.C. Gonzalez-Garcia {\it et al.}, Phys.Rev. {\bf D63} (2001) 033005;
M.C. Gonzalez-Garcia and C. Pe\~na-Garay, Nucl. Phys. Proc. Suppl. 
{\bf 91} (2000) 80 [hep-ph/0009041]

\bibitem{chooz} M. Apollonio {\it et al.}, Phys.Lett. {\bf B466} (1999) 415.

\bibitem{MSW}
L. Wolfenstein, Phys. Rev. {\bf D17} (1978) 2369;
S. Mikheyev and A. Yu. Smirnov, Sov. J. Nucl. Phys. {\bf 42} (1985) 913.

\bibitem{SNO3}
Q.~R.~Ahmad {\it et al.}  [SNO Collaboration],
nucl-ex/0106015.

\bibitem{BPB}
J.~N.~Bahcall, M.~H.~Pinsonneault and S.~Basu,
ApJ {\bf 555} (2001) 990 [astro-ph/0010346]

\bibitem{recent}
G.~L.~Fogli, E.~Lisi, D.~Montanino and A.~Palazzo,
hep-ph/0106247;
John N. Bahcall, M. C. Gonzalez-Garcia, Carlos Pena-Garay,
hep-ph/0106258.



\bibitem{SelModels} For a non-exhaustive list of recent reviews, 
see for example: 
G. Altarelli and F. Feruglio, Phys. Rep. {\bf 320} (1999) 295;
R. N. Mohapatra, hep-ph/0008232; 
J.W.F. Valle, hep-ph/9911224;
S.M. Bilenky, C. Giunti, W. Grimus, Prog.Part.Nucl.Phys. 43 (1999) 1-86;
S.F. King, hep-ph/0105261.

\bibitem{seesaw}
M. Gell-Mann, P. Ramond and R. Slansky in Sanibel Talk,
CALT-68-709, Feb 1979, and in {\it Supergravity} (North Holland,
Amsterdam 1979);
T. Yanagida in {\it Proc. of the Workshop on Unified Theory and
Baryon Number of the Universe}, KEK, Japan, 1979.


\bibitem{NtrlDecay} W. Porod {\it et al.}, Phys. Rev. {\bf D63} (2001) 115004

\bibitem{Raidal} E. Ma {\it et al.}, hep-ph/0012101

\bibitem{flavour}
C. D. Froggatt and H. B. Nilsen, Nucl. Phys. {\ B147} (1979)
277; L. Ibanez and G.G. Ross, Phys. Lett. B332 (1994) 100;
P. Binetruy and P. Ramond, Phys. Lett. B350 (1995) 49.

\bibitem{Steve0} 
S.F. King, Nucl. Phys. {\bf B562} (1999) 57.

\bibitem{Steve} S.F. King, Nucl. Phys. {\bf B576} (2000) 85.

\bibitem{KingNimai} S.F. King and N.N. Singh, 
Nucl. Phys. {\bf B591} (2000) 3

\bibitem{HMW}
L.~Hall, H.~Murayama and N.~Weiner,
Phys.\ Rev.\ Lett.\ {\bf 84}, 2572 (2000)
[hep-ph/9911341].

\bibitem{nf}
C.Albright et al, hep-ex/0008064.

\end{thebibliography}
\end{document}